\documentclass[fleqn,10pt]{wlscirep}
\usepackage[utf8]{inputenc}
\usepackage[T1]{fontenc}
\usepackage{epsfig}
\usepackage{epstopdf}
\usepackage{nicefrac}
\usepackage{graphics}
\usepackage{float}
\usepackage{amssymb}
\usepackage{graphicx,subcaption}
\usepackage{mathtools}
\usepackage{xcolor}
\usepackage{amsfonts}
\IfFileExists{dsfont.sty}{\usepackage{dsfont}}{\providecommand{\mathds}[1]{\mathbf{##1}}}
\usepackage{bm}
\usepackage{verbatim}
\usepackage{placeins}

\hypersetup{colorlinks=true,allcolors=blue,citecolor=blue,linkcolor=blue,urlcolor=blue}

\graphicspath{{./}{figs/}{Figure/}{Figures/}}

\DeclareMathOperator{\sech}{sech}
\DeclareMathOperator{\Tr}{Tr}
\providecommand{\ket}{}\renewcommand{\ket}[1]{\left|#1\right\rangle}
\providecommand{\bra}{}\renewcommand{\bra}[1]{\left\langle#1\right|}
\providecommand{\braket}{}\renewcommand{\braket}[2]{\left\langle#1\middle|#2\right\rangle}
\newcommand{\ba}{\bm a}
\newcommand{\bn}{\bm n}

\title{Axis-selective pulse control for thermometry and coupling-strength sensing of a spin bath with a central-spin probe}

\author[1]{Huzaifa Amjad}
\author[1]{Ayesha Kibria}
\author[1,2,*]{Ali Raza Mirza}
\affil[1]{Department of Physics, Quaid-i-Azam University, Islamabad 45320, Pakistan}
\affil[2]{National Centre for Physics, Islamabad 45320, Pakistan}
\affil[*]{ali.mirza@qau.edu.pk}


\begin{abstract}
We study how periodic $\pi$ pulses applied to a single two-level probe change the precision with which the temperature of a spin bath and the probe--bath coupling strength can be estimated, with and without the probe--bath correlations present in the joint thermal equilibrium state. The probe is a biased, tunnelling central spin coupled to $N$ bath spins with an optional nearest-neighbour interaction $\chi$. Writing the evolution over one pulse period as an SU(2) rotation, we obtain the probe dynamics in closed form for any number of instantaneous pulses about any axis. The choice of pulse axis is decisive. Conventional decoupling pulses about $x$ or $y$ refocus the static bath and erase the information, whereas pulses about the bias axis $z$ suppress the probe's tunnelling and leave the information-carrying coupling intact, turning the dynamics into pure dephasing. In this limit the optimal uncorrelated temperature quantum Fisher information (QFI) no longer depends on the coupling or the tunnelling. The pulses raise the QFI for both parameters, most strongly at weak coupling and large tunnelling, while initial correlations help most at intermediate and strong coupling, so the two act as complementary resources. A measurement in the eigenbasis of the symmetric logarithmic derivative saturates the QFI with and without pulses.
\end{abstract}

\begin{document}
\flushbottom
\maketitle

\section{Introduction}
\label{sec:intro}

Every quantum system interacts with its surroundings, and this interaction leads to decoherence~\cite{BreuerPetruccione2007,Weiss2008}. To understand and eventually control the decoherence one has to learn the parameters that characterise the environment, such as its temperature and its coupling strength to the system. A powerful way to do so is to use a quantum probe: a small, controllable quantum system that interacts with the environment and is then measured~\cite{Benedetti2018,Razavian2019,Gebbia2020,WuShi2020,Tamascelli2020,Correa2015,Mehboudi2019,DegenReinhardCappellaro2017}. The ultimate precision of such an estimate is set by the quantum Fisher information (QFI) through the quantum Cram\'er--Rao bound, $\delta x\geq1/\sqrt{\nu F(x)}$ for $\nu$ repetitions~\cite{Helstrom1976,BraunsteinCaves1994,Paris2009}. Any strategy that increases the QFI therefore improves the probe. Two such strategies have been studied in our earlier work. The first exploits the correlations that build up between the probe and the environment before the probe state is prepared~\cite{RomeroPaz1997,Lutz2003,BanerjeeGhosh2003,SmirneBreuerPiiloVacchini2010,DijkstraTanimura2010,MorozovMatheySRopke2012,ChaudhryGong2013a,ChaudhryGong2013b,ChaudhryGong2013c,BuserCerrilloSchallerCao2017,deVegaAlonso2017,mirza2021master,MajeedChaudhry2019,mirza2024role}. If the probe and environment first reach their joint equilibrium state and the probe is only then prepared by a projective measurement, the environment keeps a memory of the interaction and this memory carries extra information about the environment~\cite{AtherChaudhry2021,ZhangGong2024,mirza2023improving}. For a central spin probe coupled to a spin bath we showed that these initial correlations can noticeably improve the estimation of the bath temperature and of the probe--bath coupling strength~\cite{mirza2024impact}, and that the improvement depends on how the probe state is prepared~\cite{mirza2025effect}. The second strategy uses more than one probe, and two-qubit probes estimate environment parameters more precisely than single qubits~\cite{mirza2024improving,abbas2025sensing,khan2026improving}.

A third, independent resource is quantum control. Sequences of $\pi$ pulses, from the Hahn echo and the Carr--Purcell--Meiboom--Gill sequence~\cite{Hahn1950,CarrPurcell1954,MeiboomGill1958} to periodic and optimised dynamical decoupling~\cite{ViolaLloyd1998,ViolaKnillLloyd1999,KhodjastehLidar2005,Uhrig2007,SouzaAlvarezSuter2012}, are the standard tools for protecting qubits from their environment~\cite{deLangeWangRisteDobrovitskiHanson2010,PokharelAnandFortmanLidar2018,Mukherjee2013,Addis2015}. They are also the basis of noise spectroscopy~\cite{NorrisPazSilvaViola2016,DegenReinhardCappellaro2017}, they can restore favourable scaling in noisy metrology~\cite{Sekatski2016}, and periodic modulation of a probe can improve low-temperature thermometry~\cite{Mukherjee2019thermo}. Ather and Chaudhry showed for a harmonic oscillator environment that combining control pulses with initial correlations increases the QFI considerably~\cite{AtherChaudhry2021}. We suggested in our earlier study of this system~\cite{mirza2024impact} that pulses could lengthen the coherence time of a central spin in a spin bath and so improve the estimates. Whether they actually do so is not obvious. A spin bath of the type considered there is static: each bath configuration simply shifts the probe's energy bias, and the probe dephases because it averages over these shifts~\cite{Zurek1982,ProkofevStamp2000,CucchiettiPazZurek2005}. An echo removes exactly this spread, but the spread is also what carries the information about the bath. Because the probe also tunnels, its dynamics is not of pure dephasing type, and the effect of the pulses depends on which part of the probe Hamiltonian they reverse.

In this paper we answer this question exactly. We add periodic instantaneous $\pi$ pulses about an arbitrary axis to the spin--spin model studied previously~\cite{mirza2024impact} and solve the probe dynamics in closed form for both the uncorrelated (product) and the correlated (projected thermal) initial state. With the pulses switched off the results reduce exactly to our earlier pulse-free results~\cite{mirza2024impact}, which we use as a benchmark. Our main findings are: (i) conventional decoupling pulses, about $x$ or $y$, refocus the static bath and destroy the information about it; (ii) pulses about the bias axis $z$ suppress the probe's tunnelling, in the spirit of coherent destruction of tunnelling~\cite{Grossmann1991}, while preserving the probe--bath coupling, and they increase the QFI for both the temperature and the coupling strength; (iii) in the fast-pulse limit the QFI follows from a closed-form expression, and the optimal uncorrelated temperature QFI becomes independent of the coupling strength and of the tunnelling amplitude; (iv) control pulses and initial correlations are complementary, the former dominating at weak and the latter at strong coupling. We also study the effect of the bath size and of interactions between the bath spins, and we derive the classical Fisher information (CFI) of projective probe measurements to identify a measurement that reaches the QFI. The paper is organised as follows. Section~\ref{sec:prep} describes the preparation of the initial state. Section~\ref{sec:model} presents the model and its exact dynamics with and without pulses, with details in the appendices. Section~\ref{sec:estimation} derives the QFI and presents the estimation of the temperature and of the coupling strength, the role of the bath size and of inter-spin interactions, and the optimal measurement. Section~\ref{sec:conclusion} concludes.

\section{Preparation of the initial state}
\label{sec:prep}

\begin{figure}[!htbp]
\centering
\includegraphics[width=0.95\linewidth,trim=0 85bp 0 0,clip]{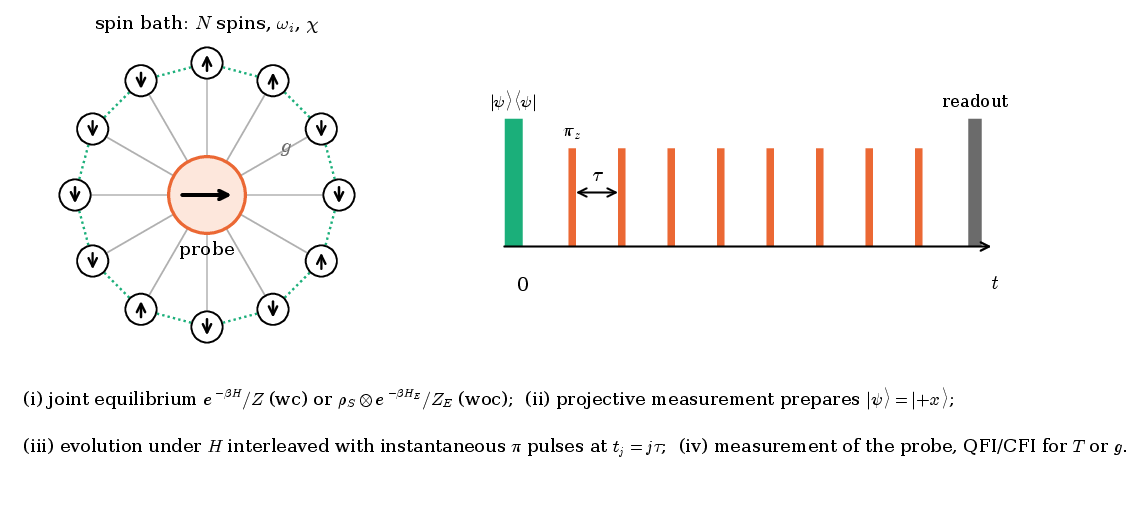}
\caption{The central spin (probe) is coupled with strength $g$ to $N$ bath spins with level splittings $\omega_i$ and nearest-neighbour interaction $\chi$. The probe state $\ket{+x}$ is prepared by a projective measurement, either from the joint equilibrium state $e^{-\beta H}/Z$ (wc) or with the bath in its own thermal state $e^{-\beta H_E}/Z_E$ (woc). The probe then evolves under the total Hamiltonian $H$ while instantaneous $\pi$ pulses are applied at the times $t_j=j\tau$. Finally the probe is measured, and the QFI or CFI is evaluated for $T$ or $g$.}
\label{fig:schematic}
\end{figure}

The usual choice of initial probe--environment state is a product state. Preparing the probe in $\ket\psi$ by a selective projective measurement at $t=0$ gives
\begin{equation}
\rho^{u}_{\rm tot}(0)=\ket\psi\bra\psi\otimes\frac{e^{-\beta H_E}}{Z_E},\qquad Z_E=\Tr_E\bigl[e^{-\beta H_E}\bigr],
\label{eq:rhou0}
\end{equation}
with $\beta=1/T$. The superscript $u$ (``uncorrelated'', woc) indicates that any interaction between probe and environment before the preparation is ignored. Physically, however, the probe has interacted with the environment for $t<0$ and the two have reached the joint equilibrium state $e^{-\beta H}/Z$, with $H$ the total Hamiltonian. The same projective measurement applied to this state gives
\begin{equation}
\rho^{c}_{\rm tot}(0)=\ket\psi\bra\psi\otimes\frac{\bra\psi e^{-\beta H}\ket\psi}{Z},\qquad Z=\Tr_{S,E}\bigl[\ket\psi\bra\psi\, e^{-\beta H}\bigr],
\label{eq:rhoc0}
\end{equation}
where the superscript $c$ (``correlated'', wc) indicates that the probe--environment correlations present before the preparation are included~\cite{mirza2021master,mirza2024impact}. Equation~\eqref{eq:rhoc0} is again a product state, but the environment is no longer in its own equilibrium state: its state depends on $H_{SE}$ and on the prepared probe state. If the coupling is weak, Eq.~\eqref{eq:rhoc0} reduces to Eq.~\eqref{eq:rhou0}. As in the benchmark study~\cite{mirza2024impact}, we prepare the probe in the state ``up'' along $x$, $\ket\psi=\ket{+x}=(\ket{\uparrow}+\ket{\downarrow})/\sqrt2$, so that the initial Bloch vector is $\ba(0)=(1,0,0)$ for both preparations. The full protocol is sketched in Fig.~\ref{fig:schematic}: after the preparation, the probe and the environment evolve together while instantaneous $\pi$ pulses are applied to the probe at the times $t_j=j\tau$, and the probe is finally measured.

\section{The model and its dynamics}
\label{sec:model}

We use the spin--spin model of the benchmark study~\cite{mirza2024impact}, with $\hbar=k_B=1$,
\begin{align}
H_S&=\frac{\varepsilon}{2}\sigma_z+\frac{\Delta}{2}\sigma_x,\label{eq:HS}\\
H_E&=\sum_{i=1}^{N}\Bigl[\frac{\omega_i}{2}\sigma_z^{(i)}+\chi_i\,\sigma_z^{(i)}\sigma_z^{(i+1)}\Bigr],\label{eq:HE}\\
H_{SE}&=\frac12\sigma_z\otimes g\sum_{i=1}^N\sigma_z^{(i)},\label{eq:HSE}
\end{align}
where $\varepsilon$ and $\Delta$ are the energy bias and tunnelling amplitude of the probe, $\omega_i$ is the level splitting of the $i$-th bath spin, $\chi_i$ is the nearest-neighbour interaction (we take a ring, $\sigma^{(N+1)}_z\equiv\sigma_z^{(1)}$), and $g$ is the probe--bath coupling strength. We take $\omega_i=\omega$ and $\chi_i=\chi$ and express all energies in units of $\omega=1$. Unless stated otherwise, $\varepsilon=2$ and $\Delta=1$, as in the benchmark study.

The bath operators in $H_E$ and $H_{SE}$ commute, so the bath configurations $\ket{e}=\ket{e_1e_2\cdots e_N}$, $e_i=0,1$, are conserved. Each configuration is labelled by its magnetisation and bond sum,
\begin{equation}
M_e=\sum_{i=1}^N(-1)^{e_i},\qquad B_e=\sum_{i=1}^N(-1)^{e_i}(-1)^{e_{i+1}},
\label{eq:MB}
\end{equation}
has energy $E_e=\omega M_e/2+\chi B_e$, and shifts the probe bias to $\varepsilon_e=\varepsilon+gM_e$. Conditioned on $\ket e$ the probe evolves under
\begin{equation}
H_e=\frac{\varepsilon_e}{2}\sigma_z+\frac{\Delta}{2}\sigma_x=\eta_e\,\bm h_e\cdot\bm\sigma,\qquad \eta_e=\frac12\sqrt{\varepsilon_e^2+\Delta^2},\qquad \bm h_e=\frac{(\Delta,0,\varepsilon_e)}{2\eta_e}.
\label{eq:He}
\end{equation}
Because the probe dynamics depends on $\ket e$ only through $M_e$, and the weights only through $(M_e,B_e)$, all sums over the $2^N$ configurations reduce to sums over these two labels with known degeneracies (Appendix~\ref{app:counting}).

\subsection{Dynamics without pulses}

The free propagator is $U(t)=\sum_e e^{-iE_et}\,U_e(t)\ket e\bra e$ with
\begin{equation}
U_e(t)=e^{-iH_et}=\cos(\eta_et)-i\sin(\eta_et)\,\bm h_e\cdot\bm\sigma .
\label{eq:Ue}
\end{equation}
The phase $e^{-iE_et}$ drops out of the reduced state, which is
\begin{equation}
\rho^{\alpha}(t)=\sum_e p_e^{\alpha}\,U_e(t)\ket\psi\bra\psi U_e^\dagger(t)=\frac12\bigl[\mathds1+\ba^{\alpha}(t)\cdot\bm\sigma\bigr],\qquad \alpha=u,c,
\label{eq:rhot}
\end{equation}
with the configuration weights
\begin{equation}
p_e^{u}=\frac{e^{-\beta E_e}}{Z_E},\qquad
p_e^{c}=\frac{e^{-\beta E_e}Q_e}{Z},\qquad
Q_e=\bra\psi e^{-\beta H_e}\ket\psi=\cosh(\beta\eta_e)-\frac{\Delta}{2\eta_e}\sinh(\beta\eta_e).
\label{eq:weights}
\end{equation}
The initial correlations enter only through the factor $Q_e$, which reweights each configuration by how strongly it is correlated with the prepared probe state (Appendix~\ref{app:corr}). $U_e(t)$ rotates the Bloch vector by the angle $2\eta_et$ about $\bm h_e$, so
\begin{equation}
a^{\alpha}_x=\sum_ep_e^\alpha\frac{\Delta^2+\varepsilon_e^2\cos(2\eta_et)}{4\eta_e^2},\quad
a^{\alpha}_y=\sum_ep_e^\alpha\frac{\varepsilon_e}{2\eta_e}\sin(2\eta_et),\quad
a^{\alpha}_z=\sum_ep_e^\alpha\frac{\Delta\varepsilon_e}{2\eta_e^2}\sin^2(\eta_et).
\label{eq:blochfree}
\end{equation}
These are exactly the Bloch components $L^{\alpha}_{ix}(t)$ derived in the benchmark study~\cite{mirza2024impact}.

\subsection{\texorpdfstring{Dynamics with periodic $\pi$ pulses}{Dynamics with periodic pi pulses}}
\label{sec:pulses}

We now apply instantaneous $\pi$ pulses about the axis $\hat{\bm a}\in\{\hat x,\hat y,\hat z\}$ at $t_j=j\tau$, $j=1,2,\ldots$. A pulse $P_a=e^{-i\pi\sigma_a/2}=-i\sigma_a$ acts on the probe only, so it changes neither the conserved configuration $\ket e$ nor the weights $p_e^\alpha$. The pulses act after the preparation, so they affect the uncorrelated and the correlated states in the same way. At time $t=K\tau+r$, $K=\lfloor t/\tau\rfloor$, $0\le r<\tau$, the conditional propagator is
\begin{equation}
U_e^{(\tau)}(t)=U_e(r)\,\bigl[P_aU_e(\tau)\bigr]^K .
\label{eq:Upulsed}
\end{equation}
The key observation is that $\mathcal M_e\equiv P_aU_e(\tau)$ belongs to SU(2), so it is itself a rotation,
\begin{equation}
\mathcal M_e=\cos\theta_e-i\sin\theta_e\,\bn_e\cdot\bm\sigma,\qquad
\mathcal M_e^K=\cos(K\theta_e)-i\sin(K\theta_e)\,\bn_e\cdot\bm\sigma .
\label{eq:Floquet}
\end{equation}
With $c_e=\cos(\eta_e\tau)$ and $s_e=\sin(\eta_e\tau)$, multiplying out $P_aU_e(\tau)$ gives (Appendix~\ref{app:pulses})
\begin{align}
\pi_x:&\quad \cos\theta_e=-\frac{\Delta}{2\eta_e}s_e,\qquad \sin\theta_e\,\bn_e=\Bigl(c_e,\,-\frac{\varepsilon_e}{2\eta_e}s_e,\,0\Bigr),\label{eq:px}\\
\pi_y:&\quad \cos\theta_e=0,\qquad\qquad\ \ \ \bn_e=\Bigl(\frac{\varepsilon_e}{2\eta_e}s_e,\,c_e,\,-\frac{\Delta}{2\eta_e}s_e\Bigr),\label{eq:py}\\
\pi_z:&\quad \cos\theta_e=-\frac{\varepsilon_e}{2\eta_e}s_e,\qquad \sin\theta_e\,\bn_e=\Bigl(0,\,\frac{\Delta}{2\eta_e}s_e,\,c_e\Bigr).\label{eq:pz}
\end{align}
The pulsed Bloch vector therefore follows from two successive rotations of $\ba(0)=\hat x$: by $2K\theta_e$ about $\bn_e$ and then by $2\eta_er$ about $\bm h_e$,
\begin{equation}
\ba^{\alpha}(t)=\sum_ep^{\alpha}_e\,R_{\bm h_e}(2\eta_er)\,R_{\bn_e}(2K\theta_e)\,\hat x .
\label{eq:blochpulsed}
\end{equation}
At the stroboscopic times $t=K\tau$ this gives
\begin{align}
a_x^\alpha&=\sum_ep_e^\alpha\bigl[n_{x}^2+(1-n_{x}^2)\cos(2K\theta_e)\bigr],\nonumber\\
a_y^\alpha&=\sum_ep_e^\alpha\bigl[n_{x}n_{y}\{1-\cos(2K\theta_e)\}+n_{z}\sin(2K\theta_e)\bigr],\label{eq:blochstrobo}\\
a_z^\alpha&=\sum_ep_e^\alpha\bigl[n_{x}n_{z}\{1-\cos(2K\theta_e)\}-n_{y}\sin(2K\theta_e)\bigr],\nonumber
\end{align}
where $\bn=\bn_e$. Equations~\eqref{eq:blochfree} have exactly the same form, with $\bn_e\to\bm h_e$ and $K\theta_e\to\eta_et$. Equations~\eqref{eq:Upulsed}--\eqref{eq:blochstrobo} are exact for any $N$, $\chi$, $g$, $T$, $\tau$ and number of pulses.

Without pulses ($K=0$, i.e.\ $\tau>t$) Eq.~\eqref{eq:blochpulsed} reduces identically to Eq.~\eqref{eq:blochfree}, so every pulse-free result of the benchmark study~\cite{mirza2024impact} is recovered. We checked the closed form \eqref{eq:Upulsed}--\eqref{eq:pz} against a brute-force product of $2\times2$ matrix exponentials over all $2^N$ configurations (with $N=6$, $\chi=0.1$, and both preparations); the two agree to $10^{-15}$. The three axes act very differently. Since $\sigma_yH_e\sigma_y=-H_e$, a $\pi_y$ pulse reverses the whole probe Hamiltonian, $\mathcal M_e^2=-\mathds1$, and the probe returns to its initial state after every second pulse for every configuration and every $\tau$. The bath is perfectly refocused and no information about it remains. A $\pi_x$ pulse reverses the bias and the coupling, $\sigma_x(\varepsilon_e\sigma_z)\sigma_x=-\varepsilon_e\sigma_z$, but keeps the tunnelling. For $\tau\to0$, Eq.~\eqref{eq:px} gives $\theta_e\to\pi/2+\Delta\tau/2$ and $\bn_e\to\hat x$, independent of $e$: the probe precesses about $x$ with frequency $\Delta$, and since $\ket{+x}$ is an eigenstate of $\sigma_x$ it is frozen. This is conventional dynamical decoupling~\cite{ViolaLloyd1998,ViolaKnillLloyd1999}. It protects the probe state, but because the bath is static it also removes the dependence of the probe state on $g$ and $T$. A $\pi_z$ pulse does the opposite: it commutes with the bias and with the probe--bath coupling but reverses the tunnelling, $\sigma_z(\Delta\sigma_x)\sigma_z=-\Delta\sigma_x$. It is a decoupling sequence for the probe's own transverse term, the pulsed analogue of coherent destruction of tunnelling~\cite{Grossmann1991}, and it leaves the information-carrying coupling untouched.

\subsection{\texorpdfstring{Ideal $\pi_z$ control: closed-form QFI}{Ideal pi_z control: closed-form QFI}}
\label{sec:ideal}

Expanding Eq.~\eqref{eq:pz} for small $\tau$ gives (Appendix~\ref{app:pulses})
\begin{equation}
2\theta_e=\pi+\varepsilon_e\tau\Bigl(1-\frac{\Delta^2\tau^2}{24}\Bigr)+O(\tau^5),\qquad \bn_e=\hat z+O(\Delta\tau).
\label{eq:pzsmall}
\end{equation}
Up to the parameter-independent factor $(-1)^K$, the stroboscopic evolution is therefore pure dephasing under $\varepsilon_e\sigma_z/2$. The tunnelling is removed, its residual effect enters only at order $(\Delta\tau)^2$, and the expansion holds for $\eta_e\tau\ll1$. In this limit $a_z^\alpha=0$ and
\begin{equation}
a^\alpha_x(t)+ia^\alpha_y(t)=(-1)^K\sum_ep^\alpha_e\,e^{i\varepsilon_et}\equiv(-1)^K e^{-\Gamma^\alpha(t)}e^{i\Omega^\alpha(t)} .
\label{eq:coherence}
\end{equation}
For the uncorrelated state with $\chi=0$ the sum over the binomially distributed magnetisation can be done exactly. With $u\equiv\tanh(\beta\omega/2)$,
\begin{equation}
e^{-\Gamma^u}e^{i\Omega^u}=e^{i\varepsilon t}\bigl[\cos(gt)-iu\sin(gt)\bigr]^N,
\label{eq:closed}
\end{equation}
so that
\begin{equation}
\Gamma^u=-\frac N2\ln r^2,\quad \Omega^u=\varepsilon t-N\arctan\bigl[u\tan(gt)\bigr],\quad r^2\equiv\cos^2(gt)+u^2\sin^2(gt)=1-\sech^2\!\Bigl(\frac{\beta\omega}{2}\Bigr)\sin^2(gt).
\label{eq:GamOm}
\end{equation}
Equation~\eqref{eq:closed} depends on $g$ and $t$ only through $gt$, apart from the parameter-independent phase $\varepsilon t$, and it does not depend on $\Delta$ at all. This has an immediate consequence, derived below: under ideal $\pi_z$ control the optimal uncorrelated temperature QFI is the same for every coupling strength and every tunnelling amplitude.

\section{Parameter estimation}
\label{sec:estimation}

\subsection{Quantum Fisher information}
\label{sec:qfi}

The QFI of a state $\rho=\sum_n\rho_n\ket{v_n}\bra{v_n}$ with respect to a parameter $x$ is~\cite{Helstrom1976,BraunsteinCaves1994,Paris2009}
\begin{equation}
F(x)=\sum_n\frac{(\partial_x\rho_n)^2}{\rho_n}+2\sum_{n\neq m}\frac{(\rho_n-\rho_m)^2}{\rho_n+\rho_m}\bigl|\braket{v_m}{\partial_xv_n}\bigr|^2 .
\label{eq:QFIgeneral}
\end{equation}
For the qubit state $\rho=\frac12(\mathds1+\ba\cdot\bm\sigma)$, with eigenvalues $\frac12(1\pm|\ba|)$, Eq.~\eqref{eq:QFIgeneral} reduces to (Appendix~\ref{app:qfi})
\begin{equation}
F(x)=|\partial_x\ba|^2+\frac{(\ba\cdot\partial_x\ba)^2}{1-|\ba|^2}.
\label{eq:QFIbloch}
\end{equation}
Writing $a_x=e^{-\Gamma}\cos\Omega$, $a_y=e^{-\Gamma}\sin\Omega$, Eq.~\eqref{eq:QFIbloch} becomes
\begin{equation}
F(x)=\frac{\bigl(\Gamma'-a_za_z'e^{2\Gamma}\bigr)^2}{f\,(e^{2\Gamma}-f)}+\frac{(a_z'+a_z\Gamma')^2}{f}+\frac{(\Omega')^2}{e^{2\Gamma}},\qquad f=1+a_z^2e^{2\Gamma},
\label{eq:QFIp4}
\end{equation}
where the prime denotes $\partial_x$. Equation~\eqref{eq:QFIp4} coincides with Eq.~(9) of the benchmark study~\cite{mirza2024impact}. The two forms agree to $10^{-9}$ in all our calculations. Equations~\eqref{eq:QFIbloch} and \eqref{eq:QFIp4} hold with and without pulses and for both preparations; only the Bloch vector changes. For pure dephasing ($a_z=0$, $f=1$) Eq.~\eqref{eq:QFIp4} reduces to
\begin{equation}
F(x)=\frac{(\Gamma')^2}{e^{2\Gamma}-1}+e^{-2\Gamma}(\Omega')^2 .
\label{eq:QFIpd}
\end{equation}
Inserting Eq.~\eqref{eq:GamOm} gives the QFI under ideal $\pi_z$ control in closed form. For the temperature, with $\partial_Tu=-\frac{\omega}{2T^2}(1-u^2)$,
\begin{equation}
F_T^{\rm id}=\frac{N^2(\partial_Tu)^2}{r^4}\left[\frac{u^2\sin^4(gt)}{r^{-2N}-1}+r^{2N}\sin^2(gt)\cos^2(gt)\right],
\label{eq:FTideal}
\end{equation}
and for the coupling strength
\begin{equation}
F_g^{\rm id}=\frac{N^2t^2}{r^4}\left[\frac{(1-u^2)^2\sin^2(gt)\cos^2(gt)}{r^{-2N}-1}+u^2r^{2N}\right].
\label{eq:Fgideal}
\end{equation}
Equation~\eqref{eq:FTideal} depends on $g$ and $t$ only through $x=gt$. Its maximum over time, $F_T^{\rm id,opt}(T)=\max_xF_T^{\rm id}$, therefore does not depend on $g$ or $\Delta$ and is reached at $t^\ast=x^\ast/g$. At short times Eq.~\eqref{eq:Fgideal} gives $F_g^{\rm id}\simeq t^2\langle M^2\rangle=t^2[N^2u^2+N(1-u^2)]$. The term $N^2u^2$ comes from the mean field of a polarised bath and the term $N(1-u^2)$ from its fluctuations. A useful upper bound follows from the structure of Eq.~\eqref{eq:rhot}. The conditional probe states $U_e^{(\tau)}\ket\psi$ do not depend on $T$, so by the data-processing inequality no pulse sequence and no probe measurement can extract more temperature information than the classical Fisher information of the weights $p_e^\alpha$ themselves. For $\chi=0$ and the uncorrelated state this bound is $F_T\le N\omega^2\sech^2(\beta\omega/2)/(4T^4)$. At $T=0.3$ and $N=50$ it equals $205$.

We verified Eqs.~\eqref{eq:FTideal} and \eqref{eq:Fgideal} against the exact pulsed engine with $\tau=0.002$ (relative deviation below $10^{-5}$). All figures are computed from the exact expressions~\eqref{eq:weights}, \eqref{eq:blochpulsed} and \eqref{eq:QFIbloch}, with the derivatives taken by central differences (relative step $10^{-5}$). Unless stated otherwise $N=50$, $\chi=0$, $\varepsilon=2$, $\Delta=1$, $\omega=1$ and $\tau=0.05$. The optimised QFI is $F^{\rm opt}=\max_{0<t\le t_{\max}}F(t)$, where $t_{\max}$ is the interrogation window stated in each caption. In all figures, ``no pulses'' denotes the pulse-free results of the benchmark study~\cite{mirza2024impact}.

\subsection{Choice of the control axis and of the pulse spacing}
\label{sec:axis}

\begin{figure}[!htbp]
\centering
\includegraphics[width=0.95\linewidth]{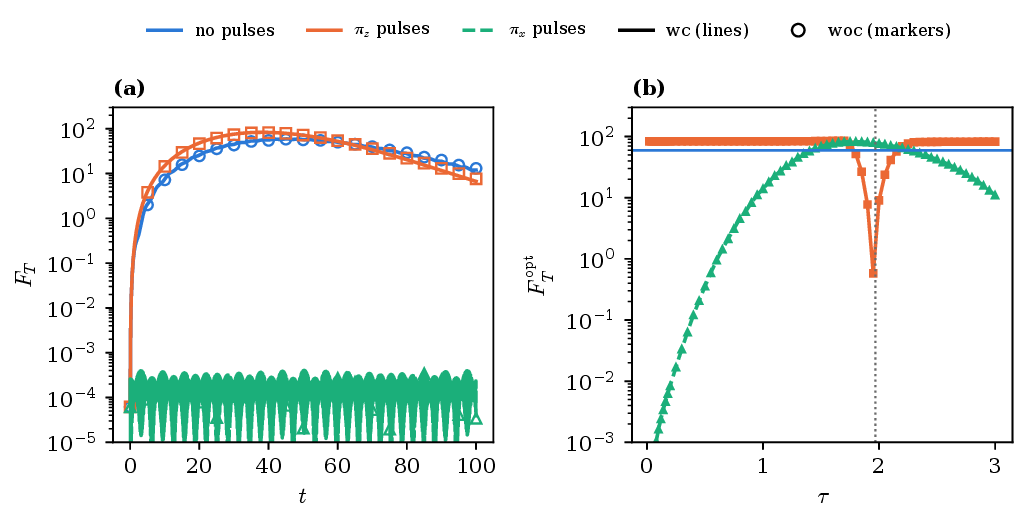}
\caption{Role of the pulse axis. (a) Temperature QFI $F_T(t)$ without pulses, with $\pi_z$ pulses and with $\pi_x$ pulses ($\tau=0.05$), for the correlated (lines) and uncorrelated (markers) preparations. (b) Optimised correlated QFI ($t_{\max}=100$) as a function of the pulse spacing $\tau$ for $\pi_z$ (squares) and $\pi_x$ (triangles) pulses; the horizontal line is the pulse-free value. The dotted vertical line marks $\tau^\ast=1.968$, where $\partial\theta_e/\partial\varepsilon_e=0$ at the mean-field bias $\bar\varepsilon=\varepsilon-gNu$. Parameters: $T=0.3$, $g=0.01$, $N=50$, $\chi=0$, $\varepsilon=2$, $\Delta=1$, $\omega=1$.}
\label{fig:axis}
\end{figure}

Figure~\ref{fig:axis} uses the parameters of Fig.~2 in the benchmark study~\cite{mirza2024impact}. Without pulses we recover the peak values reported there: $F_T^{\rm opt}=79.0$, $59.5$, $38.3$ and $23.2$ at $T=0.2$, $0.3$, $0.4$ and $0.5$, in agreement with the values reported there. At $T=0.3$, Fig.~\ref{fig:axis}(a), $\pi_z$ pulses raise the correlated peak from $59.5$ to $83.3$, a gain of $40\%$, and the peak occurs earlier ($t\approx38$ instead of $45$), because the dephasing under $\varepsilon_e\sigma_z/2$ is faster than under the tilted precession. The same $\tau=0.05$ applied about $x$ reduces the QFI to about $3\times10^{-4}$, about five orders of magnitude below the free value: the probe is frozen, as explained in Sec.~\ref{sec:pulses}.

Figure~\ref{fig:axis}(b) shows the dependence on the spacing. For $\pi_x$ pulses the QFI is strongly suppressed for small $\tau$, where the probe is frozen. It rises to a broad maximum, $F_T^{\rm opt}=82.4$ at $\tau=1.8$, close to the $\pi_z$ value, only when the spacing becomes comparable to the probe precession period, $\tau\approx1.5$--$2$, and decreases again for longer spacings. In that regime the pulses no longer decouple anything and act as a resonant modulation of the precession, not as dynamical decoupling. For $\pi_z$ pulses the QFI stays at the ideal-control value, $\simeq83$, for all $\tau\lesssim1.7$, so the advantage does not require very fast pulses. A sharp dip appears around $\tau\approx1.95$. It follows from Eq.~\eqref{eq:pz}: when $\partial\theta_e/\partial\varepsilon_e=0$ the quasi-energy of the pulsed evolution is locally independent of the bath field, so the spread of the bath field is no longer imprinted on the probe. Evaluating this condition at the mean-field bias $\bar\varepsilon=\varepsilon+g\langle M\rangle=\varepsilon-gNu=1.53$ gives $\tau^\ast=1.968$, in agreement with the numerics. In the rest of the paper we use $\pi_z$ pulses with $\tau=0.05$ and simply call them ``pulses''.

A $\pi_z$ pulse does not require a longitudinal drive: since $P_xP_y=(-i\sigma_x)(-i\sigma_y)=-i\sigma_z=P_z$, it can be applied as a back-to-back pair of conventional transverse $\pi$ pulses. For a flux-type probe it can also be applied as a short bias pulse of area $\pi$. To test the sensitivity to imperfect control we replaced every pulse by a rotation through $\pi(1+\delta)$ about an axis tilted by an angle $\phi$ from $z$, and computed the exact dynamics with the general SU(2) form of Eq.~\eqref{eq:Floquet}. At $T=0.3$ and $g=0.05$, where the pulses raise the correlated optimum from $22.1$ to $95.5$, a flip-angle error $\delta=0.1$ leaves $F_T^{\rm opt}=95.4$--$95.5$ and an axis error $\phi=0.1$ gives $94.6$--$94.8$, for both $\tau=0.05$ (2000 pulses) and $\tau=0.5$ (200 pulses). The errors are about the bias axis or close to it, so they commute with the information-carrying term $\varepsilon_e\sigma_z/2$ and only slightly spoil the averaging of the tunnelling. The protocol is therefore robust.

\subsection{Estimating the temperature of the environment}
\label{sec:temperature}

\begin{figure}[!htbp]
\centering
\includegraphics[width=\linewidth]{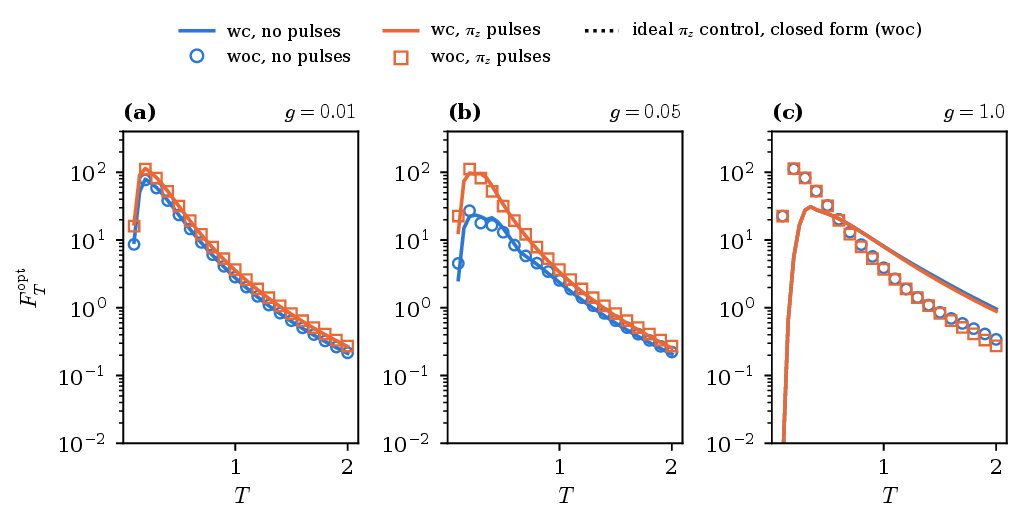}
\caption{Optimised temperature QFI as a function of $T$ for (a) $g=0.01$, (b) $g=0.05$, (c) $g=1$, without pulses (blue) and with $\pi_z$ pulses (orange), for the correlated (lines) and uncorrelated (markers) preparations. The black dotted line is the closed-form ideal-control result, Eq.~\eqref{eq:FTideal}, which is the same in all three panels. $N=50$, $\chi=0$, $t_{\max}=100$; other parameters as in Fig.~\ref{fig:axis}.}
\label{fig:temperature}
\end{figure}

Figure~\ref{fig:temperature} shows the optimised temperature QFI against $T$ at weak, intermediate and strong coupling, the regimes of Fig.~3 in the benchmark study~\cite{mirza2024impact}. In all cases the QFI peaks at a temperature of the order of $\omega$ and decreases at high $T$, where the bath distribution becomes insensitive to $T$. At weak coupling, Fig.~\ref{fig:temperature}(a), the correlated and uncorrelated results nearly coincide, as found previously~\cite{mirza2024impact}, because $Q_e$ hardly varies over the narrow range of $\varepsilon_e$. The pulses increase the QFI by a nearly constant factor of about $1.3$--$1.4$ over the whole temperature range, and the pulsed curves lie on the closed-form result \eqref{eq:FTideal}. At $g=0.05$, Fig.~\ref{fig:temperature}(b), the free QFI is lower than at $g=0.01$: the shifted biases $\varepsilon_e$ now come close to $\Delta$, the precession axis tilts strongly from configuration to configuration, and the bath field is imprinted less efficiently. With pulses the QFI returns to the universal ideal-control curve. The gain is a factor of $4$--$5$ at $T=0.2$--$0.3$ (from $22.2$ to $97.4$ at $T=0.2$ and from $22.1$ to $95.5$ at $T=0.3$, correlated). At this coupling the correlations change the pulsed QFI by up to $\pm15\%$: they increase it at $T=0.3$ and reduce it at $T=0.2$.

At strong coupling, Fig.~\ref{fig:temperature}(c), the pulsed and free curves coincide. Here $|\varepsilon_e|\gg\Delta$ for almost all configurations, so the tunnelling is already negligible and there is nothing for the pulses to remove. This regime is instead dominated by the initial correlations. At low temperature the correlated QFI is far below the uncorrelated one, because $Q_e\propto e^{\beta\eta_e}$ concentrates the weight on the configurations with the largest $|\varepsilon_e|$ and the distribution becomes insensitive to $T$. Above $T\simeq0.6$ the order is reversed, and at $T=2$ the correlated QFI is about three times larger. This crossover, also seen in the pulse-free case (Fig.~4 of the benchmark study~\cite{mirza2024impact}), persists with pulses.
\begin{figure}[!htbp]
\centering
\includegraphics[width=0.95\linewidth]{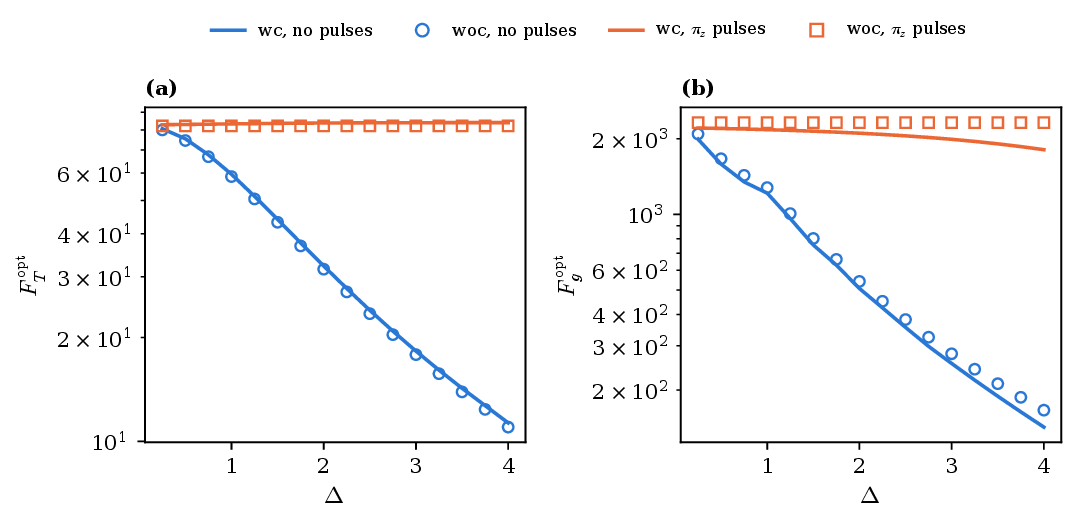}
\caption{Dependence on the probe tunnelling amplitude $\Delta$. (a) Optimised temperature QFI ($T=0.3$, $g=0.01$, $t_{\max}=100$). (b) Optimised coupling QFI ($T=1$, $g=0.05$, $t_{\max}=50$). Without pulses (blue), with $\pi_z$ pulses (orange); correlated (lines) and uncorrelated (markers). $N=50$, $\chi=0$, $\varepsilon=2$.}
\label{fig:Delta}
\end{figure}
The mechanism is confirmed by Fig.~\ref{fig:Delta}, where the tunnelling amplitude is varied. Without pulses the optimal temperature QFI falls steeply with $\Delta$, from $75$ at $\Delta=0.5$ to $11$ at $\Delta=4$, because a larger tunnelling tilts the precession axis further away from $z$ and reduces the visibility of the bath-induced phase. With pulses it stays flat at the ideal-control value, so the gain grows with $\Delta$, from $1.1$ at $\Delta=0.5$ to $7.4$ at $\Delta=4$. The coupling-strength QFI, Fig.~\ref{fig:Delta}(b), shows the same behaviour, with an even larger gain: from $1.4$ at $\Delta=0.5$ to $12.7$ at $\Delta=4$. The control is therefore most useful precisely for probes whose tunnelling is comparable to or larger than their bias.
\begin{figure}[!htbp]
\centering
\includegraphics[width=0.95\linewidth]{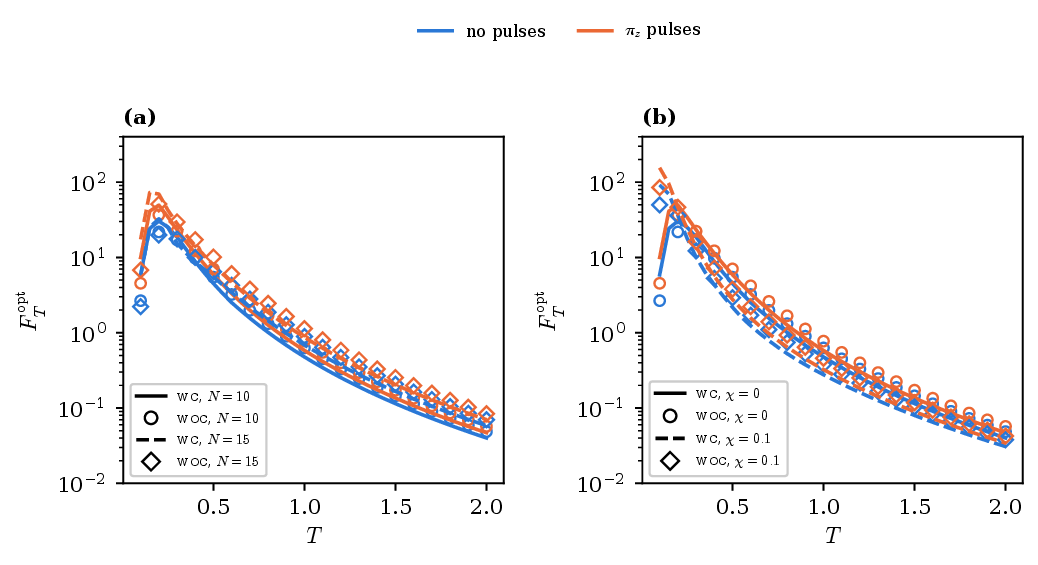}
\caption{Effect of (a) the number of bath spins, $N=10$ (solid lines, circles) and $N=15$ (dashed lines, diamonds), with $\chi=0$, and (b) the inter-spin interaction, $\chi=0$ (solid lines, circles) and $\chi=0.1$ (dashed lines, diamonds), with $N=10$, on the optimised temperature QFI. Blue: no pulses; orange: $\pi_z$ pulses. Lines: correlated; markers: uncorrelated. $g=0.1$, $t_{\max}=100$, other parameters as in Fig.~\ref{fig:axis}.}
\label{fig:Nchi_T}
\end{figure}
Figure~\ref{fig:Nchi_T} repeats the analysis of Fig.~4 in the benchmark study~\cite{mirza2024impact} for the bath size and the inter-spin interaction, at the intermediate coupling $g=0.1$ where both the pulses and the correlations matter. In panel (a), the larger bath gives a larger QFI for $T\gtrsim0.4$, both with and without pulses, in agreement with the benchmark results~\cite{mirza2024impact}: a larger bath imprints a broader distribution of fields on the probe. At lower temperature the ordering depends on the preparation. The pulses increase the QFI at every temperature and for both bath sizes. At the peak ($T\approx0.2$) the correlated QFI grows from $30.4$ to $49.5$ ($N=10$) and from $32.6$ to $69.6$ ($N=15$). The correlated preparation gives the larger QFI below $T\approx0.4$ and the smaller one above, a crossover that survives the control. In panel (b), the antiferromagnetic interaction $\chi=0.1$ creates low-lying excitations whose population changes rapidly at low temperature. The QFI at $T=0.1$ therefore grows by more than an order of magnitude, to $91.5$ (free) and $155$ (pulsed) for the correlated preparation. At higher temperature the interaction reduces the QFI, as reported previously~\cite{mirza2024impact}. In both cases the pulses act on top of the effect of $\chi$ and do not change its sign.

\subsection{Estimating the probe--environment coupling strength}
\label{sec:coupling}

\begin{figure}[!htbp]
\centering
\includegraphics[width=0.95\linewidth]{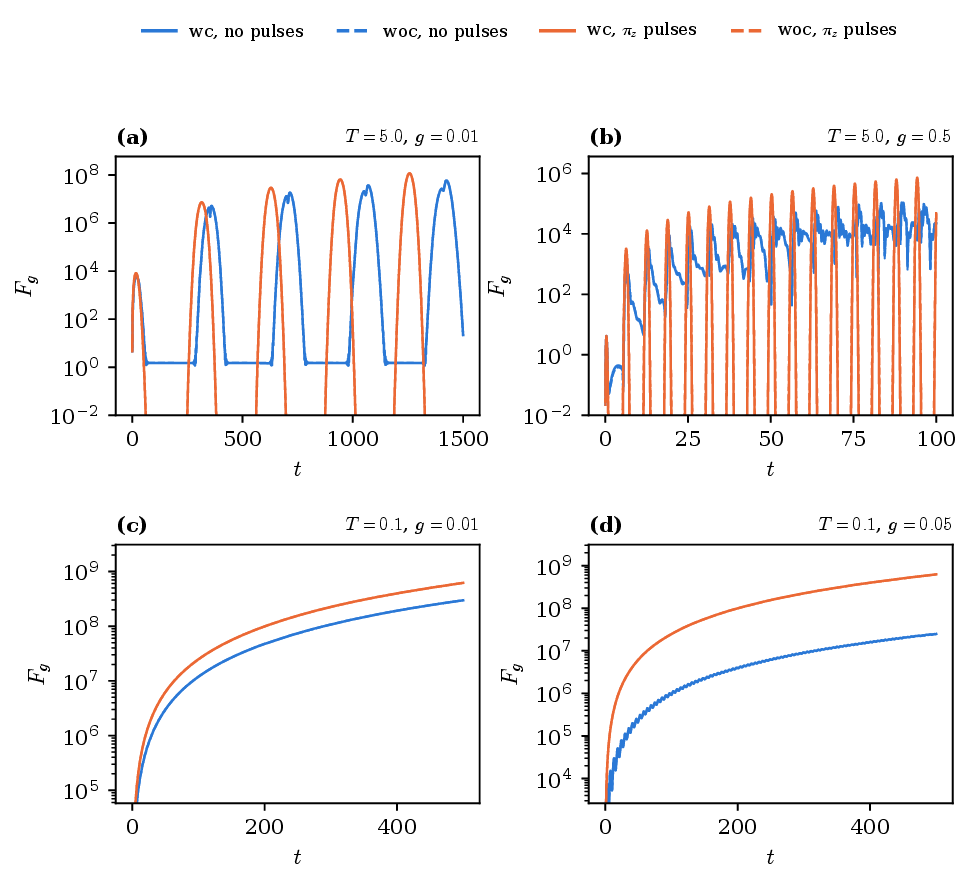}
\caption{Coupling-strength QFI $F_g$ as a function of time, without pulses (blue) and with $\pi_z$ pulses (orange), for the correlated (solid) and uncorrelated (dashed) preparations: (a) $T=5$, $g=0.01$; (b) $T=5$, $g=0.5$; (c) $T=0.1$, $g=0.01$; (d) $T=0.1$, $g=0.05$. These are the parameters of Figs.~5 and 6 in the benchmark study~\cite{mirza2024impact}. $N=50$, $\chi=0$.}
\label{fig:gtime}
\end{figure}

To estimate the coupling strength we differentiate with respect to $g$. Figure~\ref{fig:gtime} shows $F_g(t)$ for the parameters of Figs.~5 and 6 in the benchmark study~\cite{mirza2024impact}. As found there, the QFI keeps growing with time instead of reaching a single optimum: $g$ multiplies the time in the accumulated phase, so $\partial_g\ba$ grows linearly with $t$ (Eq.~\eqref{eq:Fgideal}). At high temperature, panels (a) and (b), the probe dephases between revivals. In panel (a) the free QFI shows the recurrences of the finite bath, with a period close to $2\pi/[g\,\partial(2\eta_e)/\partial M_e]=\pi/(g\,\varepsilon/2\eta)\approx350$. The pulses turn these into exact pure-dephasing revivals with period $\pi/g\approx314$ and raise the envelope: the maximum over $t\le1500$ grows from $6.0\times10^7$ to $1.2\times10^8$. At $g=0.5$, panel (b), the free QFI saturates near $10^4$--$10^5$, while the pulsed revivals reach $7\times10^5$ by $t=100$, about seven times more. At low temperature, panels (c) and (d), the bath is strongly polarised ($u\simeq1$), the coherence hardly decays, and $F_g$ grows as $t^2$ with the mean-field prefactor $N^2u^2$. The pulses give a factor of $2$ at $g=0.01$ and a factor of $25$ at $g=0.05$ ($2.5\times10^7\to6.2\times10^8$ at $t=500$). At $g=0.05$ the polarised bath shifts the bias to $\bar\varepsilon=\varepsilon-gN\approx-0.5$, comparable to $\Delta$, so the free probe is dominated by tunnelling.

\begin{figure}[!htbp]
\centering
\includegraphics[width=0.95\linewidth]{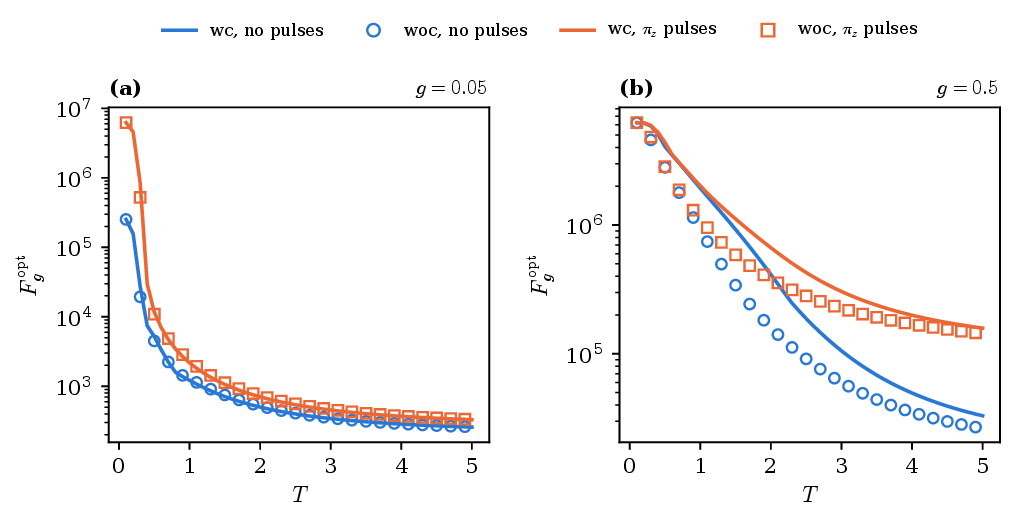}
\caption{Optimised coupling QFI ($t_{\max}=50$) as a function of temperature for (a) $g=0.05$ and (b) $g=0.5$. Blue: no pulses; orange: $\pi_z$ pulses; lines: correlated; markers: uncorrelated. $N=50$, $\chi=0$.}
\label{fig:gT}
\end{figure}

For these parameters the correlated and uncorrelated QFIs nearly coincide in panels (a), (c) and (d) (relative difference below $1\%$), both with and without pulses. In panel (b) the correlated QFI is $5$--$15\%$ larger. This agrees with Fig.~5 of the earlier work~\cite{mirza2024impact} at $T=5$. At $T=0.1$ we do not find the large correlated advantage reported in the earlier work~\cite{mirza2024impact} (Fig.~6(b)): the bath is almost fully polarised, so the $g$-dependence of the weights $Q_e$ adds very little information. As Fig.~\ref{fig:gT} shows, the initial correlations become important for the coupling estimation at larger $g$ and moderate temperatures.

Figure~\ref{fig:gT} shows the optimised coupling QFI within $t\le50$ as a function of temperature. At $g=0.05$, panel (a), the pulses give the dominant improvement, especially at low temperature, where the gain exceeds an order of magnitude, while the correlations change the QFI by less than $30\%$. At $g=0.5$, panel (b), the roles are reversed at intermediate temperatures. The correlated preparation exceeds the uncorrelated one by a factor of $2$--$2.6$ around $T=1$--$2$ without pulses. At low temperature the polarised bath shifts the bias far from zero and the pulses do nothing. At higher temperature the bath populates configurations with $|\varepsilon_e|\lesssim\Delta$, and the pulses add a factor of $1.6$ at $T=2$ and almost $5$ at $T=5$. The largest QFI is always obtained by combining the two: the pulses remove the loss of sensitivity caused by the tunnelling, and the correlations add the information stored in the weights $Q_e$. The two resources are complementary.

\begin{figure}[H]
\centering
\includegraphics[width=0.95\linewidth]{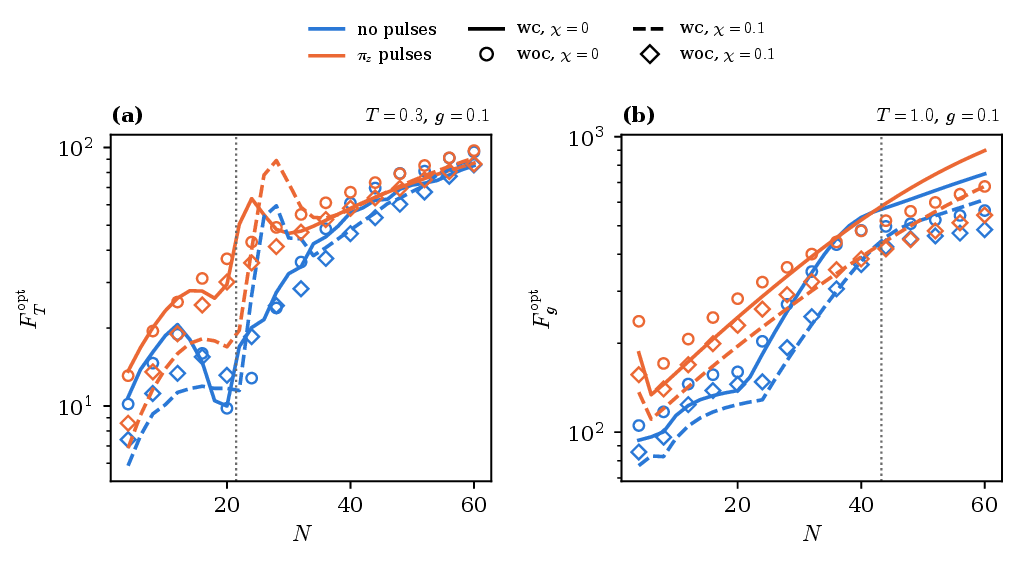}
\caption{Optimised (a) temperature QFI ($T=0.3$, $t_{\max}=100$) and (b) coupling QFI ($T=1$, $t_{\max}=20$) as a function of the number of bath spins $N$ for $\chi=0$ (solid lines, circles) and $\chi=0.1$ (dashed lines, diamonds). Blue: no pulses; orange: $\pi_z$ pulses; lines: correlated; markers: uncorrelated. The dotted vertical line marks $N^\ast=\varepsilon/[g\tanh(\beta\omega/2)]$, where the mean bath field cancels the probe bias. $g=0.1$.}
\label{fig:Nscaling}
\end{figure}

Figure~\ref{fig:Nscaling} shows how the optimal QFI depends on the bath size. Both QFIs grow with $N$ overall, consistent with $F_g\propto\langle M^2\rangle$ and with the broader field distribution of a larger bath. The growth is not monotonic, however. Its structure is organised around the bath size $N^\ast=\varepsilon/[g\tanh(\beta\omega/2)]$ at which the mean bath field $g\langle M\rangle=-gNu$ cancels the probe bias ($N^\ast\approx21.5$ at $T=0.3$ and $N^\ast\approx43$ at $T=1$). Near $N^\ast$ the effective bias of the dominant configurations is small compared with $\Delta$, the free probe is dominated by tunnelling, and the free QFI dips (for example, $9.99$ at $N=20$ against $14.8$ at $N=16$, correlated). The $\pi_z$ pulses remove the tunnelling and therefore help most near $N^\ast$, by a factor of three at $N=20$ for the temperature. Far from $N^\ast$ the pulsed and free results approach each other. The inter-spin interaction $\chi=0.1$ lowers the coupling QFI at all $N$. For the temperature QFI it mostly lowers the values at small $N$, but it also shifts the bias-compensation structure and can raise the QFI locally (for example, $59.5$ against $27.5$ at $N=28$ without pulses). The overall picture does not change. Choosing the bath size, or equivalently the probe bias, close to $N^\ast$ is thus where control fields are most valuable.

\subsection{Optimal measurement and classical Fisher information}
\label{sec:cfi}

The QFI is an upper bound that is reached only by an optimal measurement. For a projective measurement of the probe along the unit vector $\bn$, with projectors $\Pi_\pm=\frac12(\mathds1\pm\bn\cdot\bm\sigma)$, the outcome probabilities are $p_\pm=\frac12(1\pm\bn\cdot\ba)$ and the classical Fisher information is
\begin{equation}
F_C(x;\bn)=\sum_\pm\frac{(\partial_xp_\pm)^2}{p_\pm}=\frac{(\bn\cdot\partial_x\ba)^2}{1-(\bn\cdot\ba)^2}.
\label{eq:CFI}
\end{equation}
The symmetric logarithmic derivative (SLD) $L$, defined by $\partial_x\rho=\frac12(L\rho+\rho L)$, is for a qubit $L=-\frac{\ba\cdot\partial_x\ba}{1-|\ba|^2}\mathds1+\bm l\cdot\bm\sigma$ with
\begin{equation}
\bm l=\partial_x\ba+\frac{\ba\cdot\partial_x\ba}{1-|\ba|^2}\,\ba
\label{eq:SLD}
\end{equation}
(Appendix~\ref{app:qfi}). Measuring along $\bn_{\rm opt}=\bm l/|\bm l|$, i.e.\ in the eigenbasis of the SLD, gives $\bn_{\rm opt}\cdot\partial_x\ba=F/|\bm l|$ and $\bn_{\rm opt}\cdot\ba=(\ba\cdot\partial_x\ba)/[(1-|\ba|^2)|\bm l|]$. Substituting these into Eq.~\eqref{eq:CFI} gives $F_C(x;\bn_{\rm opt})=F(x)$: the SLD measurement saturates the QFI with and without pulses and for both preparations. Under ideal $\pi_z$ control $a_z=0$ and $\bm l$ lies in the $xy$ plane, so the optimal measurement is a $\sigma_\phi=\cos\phi\,\sigma_x+\sin\phi\,\sigma_y$ measurement, the standard Ramsey readout with a phase $\phi$ that depends on the estimated parameter and on $t$. A fixed $\sigma_x$ measurement gives $F_C(x;\hat x)=(\partial_xa_x)^2/(1-a_x^2)$.

\begin{figure}[!htbp]
\centering
\includegraphics[width=0.95\linewidth]{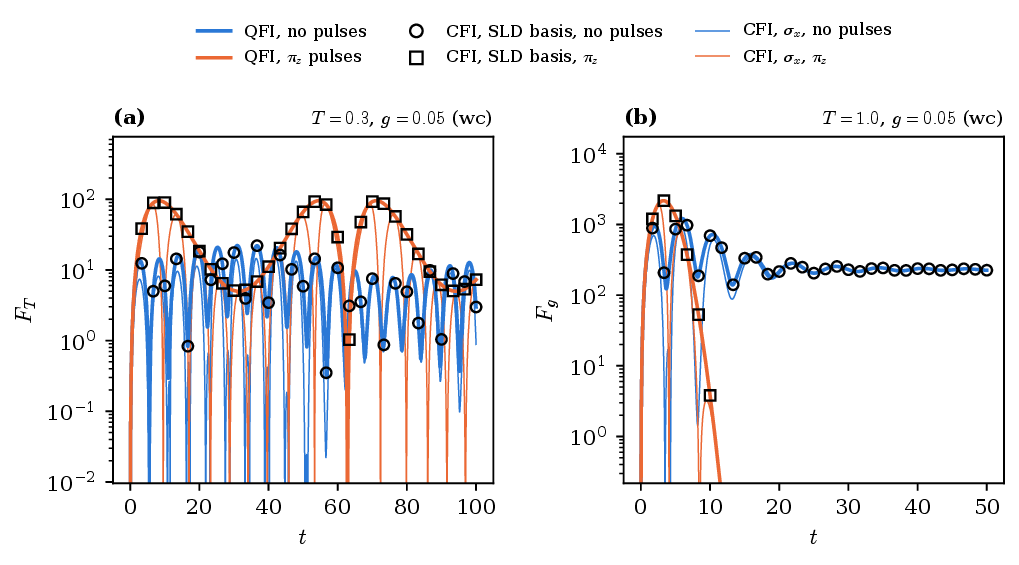}
\caption{Classical Fisher information for (a) the temperature ($T=0.3$, $g=0.05$) and (b) the coupling strength ($T=1$, $g=0.05$), correlated preparation. Thick lines: QFI without (blue) and with (orange) $\pi_z$ pulses. Open markers: CFI of the SLD-eigenbasis measurement, Eq.~\eqref{eq:CFI} with $\bn=\bn_{\rm opt}$. Thin lines: CFI of a fixed $\sigma_x$ measurement. $N=50$, $\chi=0$.}
\label{fig:cfi}
\end{figure}

Figure~\ref{fig:cfi} compares these quantities. In both panels the CFI of the SLD measurement coincides with the QFI at all times (relative deviation below $10^{-10}$ wherever the QFI is non-negligible), with and without pulses. The CFI of the fixed $\sigma_x$ measurement oscillates at the precession frequency. Its maximum reaches $90\%$ (pulses) and $66\%$ (no pulses) of the maximal QFI in panel (a), and $88\%$ and $78\%$ in panel (b). Under $\pi_z$ control the state stays in the $xy$ plane, so a fixed in-plane readout timed to its maxima is already close to optimal, and a $\sigma_\phi$ readout with the appropriate phase reaches the QFI. Without pulses the state leaves the plane and the fixed $\sigma_x$ readout loses more. In panel (a) the pulsed QFI shows periodic revivals with period $\pi/g\approx63$, each reaching $F_T\approx95$, whereas the free QFI stays below $23$. In panel (b) the pulses double the peak QFI ($2.2\times10^3$ against $1.2\times10^3$), but the advantage is limited to short times. Under $\pi_z$ control the probe state lies in the $xy$ plane, and once the coherence has dephased the information is lost until the next revival. Without pulses the population $a_z$ retains a steady $g$-dependence, which keeps $F_g\approx2\times10^2$ at long times. If the interrogation time is limited, the pulsed protocol is therefore preferable for the coupling strength. For long times a free evolution read out in the $\sigma_z$ basis can be competitive.

\FloatBarrier
\section{Conclusion}
\label{sec:conclusion}

We have shown how periodic $\pi$ pulses on a single two-level probe change the estimation of the temperature of a spin bath and of the probe--bath coupling strength, with and without the correlations in the joint probe--bath equilibrium state. Because the bath configurations are conserved, the evolution during one pulse period is an SU(2) rotation for every configuration, and we have obtained the probe dynamics in closed form for any pulse axis, spacing and number of pulses. With the pulses switched off the results reproduce our earlier benchmark results~\cite{mirza2024impact} exactly. The central result is that the pulses help only if they decouple the right term. Conventional dynamical decoupling, with pulses about $x$ or $y$, reverses the probe--bath coupling. For a static spin bath this refocuses exactly the field spread that carries the information, and the QFI collapses (by about five orders of magnitude for $\tau=0.05$). Pulses about the bias axis $z$ instead remove the tunnelling, leave the coupling untouched, and convert the dynamics into pure dephasing. They increase the optimal temperature QFI by a factor of $1.3$--$1.4$ at weak coupling, $4$--$5$ at $g=0.05$, and up to $7.4$ for $\Delta=4$, and the coupling QFI by factors of $2$ to $25$. In the fast-pulse limit we obtained the QFI in closed form. The optimal uncorrelated temperature QFI then becomes independent of the coupling strength and of the tunnelling amplitude, and it is reached at the time $t^\ast\propto1/g$. The pulses do not need to be very fast: the full advantage is retained for spacings up to $\tau\approx1.7$ in units of $1/\omega$, apart from a resonance that we have located analytically.

Initial correlations and control fields are complementary resources. The pulses matter most at weak and intermediate coupling, where the tunnelling competes with the bath field. The initial correlations matter most at intermediate to strong coupling and moderate temperature, where the reweighting factor $Q_e$ varies strongly across the bath configurations. Combining the two gives the largest QFI. The bath size enters through the bias-compensation point $N^\ast$, near which the tunnelling dominates the free probe and the pulses are most effective, and a weak inter-spin interaction leaves the picture unchanged. Finally, the classical Fisher information of the SLD-eigenbasis measurement equals the QFI in every case studied, so the bounds reported here can be reached with a single-qubit projective measurement, in practice a Ramsey $\sigma_\phi$ readout. Our model treats the pulses as instantaneous and the bath as static. Flip-angle errors of $10\%$ and axis tilts of $0.1$ rad change the optimal QFI by less than $1\%$ (Sec.~\ref{sec:axis}), and the advantage persists for spacings up to $\tau\approx1.7$, so relatively few pulses are needed. Finite pulse widths would add further corrections, which can be reduced by robust sequences~\cite{SouzaAlvarezSuter2012,KhodjastehLidar2005}. Intrinsic bath dynamics, such as flip-flop terms, would make the bath noise time dependent. In that case conventional decoupling would also filter the bath spectrum, and the interplay between the two kinds of pulses discussed here is a natural direction for future work, as is the extension to two-qubit probes~\cite{mirza2024improving,abbas2025sensing,khan2026improving}.

\section*{Acknowledgements}
AR Mirza thanks the Higher Education Commission for support through the IPFP (Interim Placement for fresh PhDs) fellowship under Grant No.\ HEC/HRD/PD/IPFP/2/2025/BVI/54796.

\section*{Data availability}
All data supporting the findings of this article can be regenerated from the closed-form expressions given in the text and appendices; the Mathematica codes used for every figure are available from the authors upon reasonable request.

\appendix
\renewcommand{\theequation}{\thesection\arabic{equation}}
\setcounter{equation}{0}

\section{Bath configurations and degeneracies}
\label{app:counting}

All quantities depend on a configuration $\ket e$ only through $M_e$ and $B_e$, Eq.~\eqref{eq:MB}. Let $k$ be the number of spins with $e_i=0$, so $M=2k-N$. For $\chi=0$ only $M$ matters and the degeneracy is $\binom Nk$. For $\chi\neq0$ on a ring, a configuration with $k$ up spins arranged in $r\geq1$ blocks has $2r$ domain walls and $B=N-4r$. The number of such configurations is
\begin{equation}
D(k,r)=\frac Nr\binom{k-1}{r-1}\binom{N-k-1}{r-1},\qquad 1\le r\le\min(k,N-k),
\label{eq:Dkr}
\end{equation}
plus the two uniform configurations $k=0,N$ with $B=N$. Equation~\eqref{eq:Dkr} reduces the $2^N$-term sums to $O(N^2)$ terms, and we checked it against direct enumeration for $N\le10$. All sums in the main text are of the form $\sum_e(\cdots)=\sum_{k,r}D(k,r)(\cdots)$.

\section{Exact dynamics with periodic pulses}
\label{app:pulses}

With $U_e(\tau)=c_e-is_e(h_x\sigma_x+h_z\sigma_z)$, $h_x=\Delta/2\eta_e$, $h_z=\varepsilon_e/2\eta_e$, and the identities $\sigma_a\sigma_b=\delta_{ab}+i\epsilon_{abc}\sigma_c$, the three pulse types give
\begin{align}
(-i\sigma_x)U_e(\tau)&=-s_eh_x-ic_e\sigma_x+is_eh_z\sigma_y,\\
(-i\sigma_y)U_e(\tau)&=-i\bigl(s_eh_z\sigma_x+c_e\sigma_y-s_eh_x\sigma_z\bigr),\\
(-i\sigma_z)U_e(\tau)&=-s_eh_z-ic_e\sigma_z-is_eh_x\sigma_y .
\end{align}
Comparing with $\cos\theta_e-i\sin\theta_e\,\bn_e\cdot\bm\sigma$ gives Eqs.~\eqref{eq:px}--\eqref{eq:pz}. In each case $\cos^2\theta_e+\sin^2\theta_e|\bn_e|^2=1$, as required for SU(2). The $K$-th power, Eq.~\eqref{eq:Floquet}, follows from $(\bn_e\cdot\bm\sigma)^2=\mathds1$. Equation~\eqref{eq:Upulsed} assumes that the free evolution in each period precedes the pulse; a final pulse at the readout time is a parameter-independent unitary and does not change the QFI.

A unitary $\cos\phi-i\sin\phi\,\bm m\cdot\bm\sigma$ rotates the Bloch vector by the angle $2\phi$ about $\bm m$. With Rodrigues' formula, $R_{\bm m}(\vartheta)\bm v=\bm v\cos\vartheta+(\bm m\times\bm v)\sin\vartheta+\bm m(\bm m\cdot\bm v)(1-\cos\vartheta)$, and $\bm v=\hat x$, this gives Eq.~\eqref{eq:blochstrobo}. For non-stroboscopic times the result is rotated further by $R_{\bm h_e}(2\eta_er)$, Eq.~\eqref{eq:blochpulsed}. Without pulses, $K=0$, $\bn_e\to\bm h_e$ and $K\theta_e\to\eta_et$, and Eq.~\eqref{eq:blochstrobo} becomes Eq.~\eqref{eq:blochfree}. For example, $h_x^2+(1-h_x^2)\cos(2\eta_et)=[\Delta^2+\varepsilon_e^2\cos(2\eta_et)]/4\eta_e^2$, the expression given in the benchmark study~\cite{mirza2024impact}.

For $\tau\to0$, $\cos\theta_e=-\frac{\varepsilon_e\tau}{2}\bigl(1-\frac{\eta_e^2\tau^2}{6}\bigr)+O(\tau^5)$. Using $\arccos(-y)=\frac\pi2+y+\frac{y^3}{6}+\cdots$ and $\eta_e^2=(\varepsilon_e^2+\Delta^2)/4$,
\begin{equation}
\theta_e=\frac\pi2+\frac{\varepsilon_e\tau}{2}\Bigl[1-\frac{(\varepsilon_e^2+\Delta^2)\tau^2}{24}+\frac{\varepsilon_e^2\tau^2}{24}\Bigr]=\frac\pi2+\frac{\varepsilon_e\tau}2\Bigl(1-\frac{\Delta^2\tau^2}{24}\Bigr)+O(\tau^5),
\end{equation}
which gives Eq.~\eqref{eq:pzsmall}. The axis is $\bn_e=(0,\Delta\tau/2,1)+O(\tau^2)$. The angle $2K\theta_e=K\pi+\varepsilon_et(1-\Delta^2\tau^2/24)$, with $t=K\tau$, gives a rotation about $z$ by $\varepsilon_et$ together with $K$ rotations by $\pi$ about $z$. The latter reverse $a_x$ and $a_y$ for odd $K$, which is the sign $(-1)^K$ in Eq.~\eqref{eq:coherence}. The same expansion for $\pi_x$ gives $\theta_e=\pi/2+\Delta\tau/2+O(\tau^3)$ and $\bn_e=\hat x+O(\varepsilon_e\tau)$, the frozen probe discussed in Sec.~\ref{sec:pulses}.

For the uncorrelated state with $\chi=0$, $p_e^u\propto e^{-\beta\omega M_e/2}$ and $M=2k-N$ with degeneracy $\binom Nk$. Then
\begin{equation}
\sum_ep^u_ee^{i\varepsilon_et}=e^{i\varepsilon t}\frac{\sum_k\binom Nk e^{(2k-N)(igt-\beta\omega/2)}}{\sum_k\binom Nk e^{-(2k-N)\beta\omega/2}}=e^{i\varepsilon t}\left[\frac{\cosh(igt-\beta\omega/2)}{\cosh(\beta\omega/2)}\right]^N=e^{i\varepsilon t}\bigl[\cos(gt)-iu\sin(gt)\bigr]^N,
\end{equation}
which is Eq.~\eqref{eq:closed}. Differentiating $\Gamma^u=-\frac N2\ln r^2$ and $\Omega^u=\varepsilon t-N\arctan[u\tan(gt)]$ gives
\begin{align}
\partial_T\Gamma^u&=-\frac{Nu\sin^2(gt)}{r^2}\partial_Tu, &\partial_T\Omega^u&=-\frac{N\sin(gt)\cos(gt)}{r^2}\partial_Tu,\\
\partial_g\Gamma^u&=\frac{N(1-u^2)t\sin(gt)\cos(gt)}{r^2}, &\partial_g\Omega^u&=-\frac{Nut}{r^2},
\end{align}
and Eq.~\eqref{eq:QFIpd} then yields Eqs.~\eqref{eq:FTideal} and \eqref{eq:Fgideal}.

\section{The correlated initial state}
\label{app:corr}

Since $[H_E+H_{SE},\ket e\bra e]=0$, the total Hamiltonian is block diagonal in the bath configurations, $H=\sum_e(E_e+H_e)\ket e\bra e$, with $H_e$ of Eq.~\eqref{eq:He}. Hence
\begin{equation}
\bra\psi e^{-\beta H}\ket\psi=\sum_ee^{-\beta E_e}\bra\psi e^{-\beta H_e}\ket\psi\,\ket e\bra e,\qquad
e^{-\beta H_e}=\cosh(\beta\eta_e)-\frac{\sinh(\beta\eta_e)}{\eta_e}H_e,
\end{equation}
where we used $H_e^2=\eta_e^2$. For $\ket\psi=\ket{+x}$, $\bra\psi H_e\ket\psi=\Delta/2$, which gives $Q_e$ of Eq.~\eqref{eq:weights}. Since $\Delta/2\eta_e\le1$, $Q_e>0$, so $p^c_e$ is a proper probability distribution. The correlated initial state, Eq.~\eqref{eq:rhoc0}, is therefore $\ket\psi\bra\psi\otimes\sum_ep_e^c\ket e\bra e$. It is diagonal in the conserved basis, so the reduced dynamics is that of Eq.~\eqref{eq:rhot} with $p_e^u\to p_e^c$, both with and without pulses. For the numerics we write $\ln Q_e=\beta\eta_e+\ln\bigl[\frac12(1-\frac{\Delta}{2\eta_e})+\frac12(1+\frac{\Delta}{2\eta_e})e^{-2\beta\eta_e}\bigr]$, which avoids overflow at low temperature. The uncorrelated case follows by setting $Q_e=1$.

\section{QFI and SLD of a qubit}
\label{app:qfi}

For $\rho=\frac12(\mathds1+\ba\cdot\bm\sigma)$, inserting $L=\lambda\mathds1+\bm l\cdot\bm\sigma$ into $\partial_x\rho=\frac12\{L,\rho\}$ and using $\{\bm l\cdot\bm\sigma,\ba\cdot\bm\sigma\}=2\bm l\cdot\ba$ gives $\lambda+\bm l\cdot\ba=0$ and $\lambda\ba+\bm l=\partial_x\ba$. Hence $\bm l\cdot\ba=\ba\cdot\partial_x\ba/(1-|\ba|^2)$, which gives Eq.~\eqref{eq:SLD}, and
\begin{equation}
F=\Tr\bigl[\rho L^2\bigr]=\lambda^2+|\bm l|^2+2\lambda\,\bm l\cdot\ba=|\bm l|^2-(\bm l\cdot\ba)^2=|\partial_x\ba|^2+\frac{(\ba\cdot\partial_x\ba)^2}{1-|\ba|^2},
\end{equation}
which is Eq.~\eqref{eq:QFIbloch}. Inserting $\bn_{\rm opt}=\bm l/|\bm l|$ into Eq.~\eqref{eq:CFI} and using $|\bm l|^2-(\bm l\cdot\ba)^2=F$ gives $F_C(x;\bn_{\rm opt})=F^2/[|\bm l|^2-(\ba\cdot\partial_x\ba)^2/(1-|\ba|^2)^2]=F$. For a pure state ($|\ba|\to1$) Eq.~\eqref{eq:QFIbloch} is understood as its limit; it applies only at $t=0$ here, where $F=0$.

\bibliography{references}

\end{document}